\documentclass{article}
\usepackage[top=0.8in, bottom=0.8in, left=0.8in, right=0.8in]{geometry} 
\usepackage{amssymb}
\usepackage{amsmath}
\usepackage{graphicx}
\usepackage[style=ieee]{biblatex}
\usepackage{mathtools}
\usepackage{booktabs}
\usepackage{siunitx}
\usepackage{subcaption}
\usepackage{multirow}
\newcommand\doubleplus{+\kern-1.3ex+}  
\usepackage{adjustbox}
\usepackage{tabularx}
\usepackage{hyperref}

\renewcommand{\eqref}[1]{Equation~(\ref{eq:#1})}

\newcommand{\tabref}[1]{Table~\ref{tab:#1}} 
\newcommand{\figref}[1]{Figure~\ref{fig:#1}} 

\DeclareMathOperator*{\argmin}{arg\,min}

\renewcommand{\Tilde}{\widetilde}
\renewcommand{\Hat}{\widehat}
\renewcommand{\Bar}{\overline}

\renewcommand{\vec}[1]{\ensuremath{\boldsymbol{\mathrm{#1}}}}
\newcommand{\Tvec}[1]{\ensuremath{\Tilde{\boldsymbol{\mathrm{#1}}}}}
\newcommand{\Hvec}[1]{\ensuremath{\Hat{\boldsymbol{\mathrm{#1}}}}}

\newcommand{\bvec}[1]{\ensuremath{\Bar{\boldsymbol{\mathrm{#1}}}}}

\newcommand{\Real}{{\mathbb{R}}}
\newcommand{\Complex}{{\mathbb{C}}}

\renewcommand{\bf}{\bfseries}
\newcommand{\nonsig}{\normalfont\,\textsuperscript{(ns)}}
\newcommand{\p}{\normalfont\,\textsuperscript{*}}

\newcommand{\ppp}{\normalfont\,\textsuperscript{***}}

\newcommand{\footremember}[2]{%
    \footnote{#2}
    \newcounter{#1}
    \setcounter{#1}{\value{footnote}}%
}
\newcommand{\footrecall}[1]{%
    \footnotemark[\value{#1}]%
}
\let\svthefootnote\thefootnote
\newcommand\freefootnote[1]{%
  \let\thefootnote\relax%
  \footnotetext{#1}%
  \let\thefootnote\svthefootnote%
}

\title{Multi-dynamic deep image prior for cardiac MRI}
\author{%
    Marc Vornehm\footremember{FAU}{Department Artificial Intelligence in Biomedical Engineering, Friedrich-Alexander-Universit\"at Erlangen-N\"urnberg, Erlangen, Germany}\footremember{OSU}{Department of Biomedical Engineering, The Ohio State University, Columbus, OH, USA}\footremember{SHS}{Research \& Clinical Translation, Magnetic Resonance, Siemens Healthineers AG, Erlangen, Germany}%
    \and Chong Chen\footrecall{OSU}%
    \and Muhammad Ahmad Sultan\footrecall{OSU}%
    \and Syed Murtaza Arshad\footrecall{OSU}%
    \and Yuchi Han\footremember{OSUMC}{Division of Cardiovascular Medicine, The Ohio State University, Columbus, OH, USA}%
    \and Florian Knoll\footrecall{FAU}%
    \and Rizwan Ahmad\footrecall{OSU}%
}
\date{}

\begin{document}

\maketitle
\freefootnote{}
\freefootnote{Corresponding author: Marc Vornehm (marc.vornehm@fau.de)}

\begin{center}
\large\textit{This work has been submitted to Magnetic Resonance in Medicine for publication.}
\vspace{5mm}
\end{center}

\begin{abstract}
\noindent\textbf{Purpose:}
Cardiovascular magnetic resonance imaging is a powerful diagnostic tool for assessing cardiac structure and function.
However, traditional breath-held imaging protocols pose challenges for patients with arrhythmias or limited breath-holding capacity.
This work aims to overcome these limitations by developing a reconstruction framework that enables high-quality imaging in free-breathing conditions for various dynamic cardiac MRI protocols.\\
\noindent\textbf{Methods:}
Multi-Dynamic Deep Image Prior (M-DIP), a novel unsupervised reconstruction framework for accelerated real-time cardiac MRI, is introduced.
To capture contrast or content variation, M-DIP first employs a spatial dictionary to synthesize a time-dependent intermediate image.
Then, this intermediate image is further refined using time-dependent deformation fields that model cardiac and respiratory motion.
Unlike prior DIP-based methods, M-DIP simultaneously captures physiological motion and frame-to-frame content variations, making it applicable to a wide range of dynamic applications.\\
\noindent\textbf{Results:}
We validate M-DIP using simulated MRXCAT cine phantom data as well as free-breathing real-time cine, single-shot late gadolinium enhancement (LGE), and first-pass perfusion data from clinical patients.
Comparative analyses against state-of-the-art supervised and unsupervised approaches demonstrate M-DIP's performance and versatility.
M-DIP achieved better image quality metrics on phantom data, higher reader scores on in-vivo cine and LGE data, and comparable scores on in-vivo perfusion data relative to another DIP-based approach.\\
\noindent\textbf{Conclusion:}
M-DIP enables high-quality reconstructions of real-time free-breathing cardiac MRI without requiring external training data.
Its ability to model physiological motion and content variations makes it a promising approach for various dynamic imaging applications.
\end{abstract}

\section{Introduction}
Cardiovascular magnetic resonance imaging (CMR) is a well-established modality for comprehensive structural and functional assessment of the heart.
Data acquisition in CMR is typically synchronized with an electrocardiogram (ECG) signal and performed during breath-holds.
However, this strategy is not feasible for patients who cannot hold their breath or those with arrhythmias.
Real-time free-breathing imaging offers an alternative, but it requires higher acceleration rates, leading to compromised image quality with reduced spatial and temporal resolution.
Improving the quality of real-time imaging has become an active area of research~\cite{contijoch2024future}.

Sparsity-based compressed sensing (CS) has shown promise for accelerating various MRI applications~\cite{lustig2008compressed}.
More recently, deep learning (DL)-based methods have surpassed CS in terms of image quality~\cite{yang2018dagan, jalal2021robust, knoll2020advancing}.
In particular, the end-to-end variational network and similar methods have emerged as quality benchmarks for MRI reconstruction~\cite{hammernik2018learning, aggarwal2019modl, sriram2020endend}.
However, these methods require large, fully sampled datasets for training, which are unavailable for most cardiac applications.
Even when available~\cite{chen2020ocmr}, these training datasets are typically collected under breath-holds from subjects with regular heart rhythms and within a narrow range of imaging parameters (e.\,g., resolution, contrast, and sampling pattern).
This limits their adaptability to free-breathing imaging, imaging of arrhythmic patients, and scenarios where imaging parameters differ substantially from those used in the training dataset.

Unsupervised and self-supervised learning strategies have shown great promise in applications where uncorrupted training data are unavailable.
These methods reconstruct images by leveraging redundancies in the undersampled data.
Traditional GRAPPA~\cite{griswold2002generalized} and its nonlinear extension RAKI~\cite{akcakaya2019scanspecific} exemplify self-supervised learning approaches, where relationships between neighboring k-space samples are first learned from the fully sampled autocalibration signal region and then applied to the undersampled regions of k-space.
Recently, Yaman et~al.\ introduced a calibration-free method called self-supervised learning via data undersampling (SSDU)~\cite{yaman2020self}, which divides undersampled k-space data into two subsets.
A reconstruction network is trained by taking an aliased image from one subset as input and generating a reconstructed image consistent with the other subset.
Once trained on a collection of undersampled measurements, this network can reconstruct images from unseen undersampled data.
Although this approach can be naively extended to dynamic applications, it does not exploit the structure in the temporal dimension.

Deep image prior (DIP) offers another unsupervised learning framework for solving a wide range of inverse problems~\cite{ulyanov2018deep}.
DIP is instance-specific, meaning the training is performed from scratch for each set of measurements.
In DIP, a generative network is trained to map a random code vector to an output consistent with the measurements.
A key feature of DIP is that the network structure acts as an implicit prior, obviating the need for an explicit regularization term.
A straightforward extension of DIP for dynamic problems involves training a single network to take a different random code vector at each time step and map it to an image frame consistent with the measurements.
While this approach leverages some redundancy across frames by employing a common network, it does not fully capture the structure along the temporal dimension.

Some recent extensions of DIP specifically leverage temporal structure for dynamic MRI.
Yoo et~al.\ proposed time-dependent DIP~\cite{yoo2021timedependent}, in which the code vectors are an ordered sequence of points on a specifically designed manifold.
This approach, however, requires an a priori cardiac phase estimation, which can be inferred from a proprietary ECG signal or directly from k-space data in radial or spiral acquisitions.
In a related line of work, Zou et~al.\ proposed Gen-SToRM~\cite{zou2021dynamic} for spiral real-time cine imaging.
Instead of hand-crafting the manifold in the latent domain, this approach discovers the manifold by directly optimizing the latent code vectors along with the network parameters.
Temporal regularization on the code vectors is added to enforce smoothness on the manifold.
After convergence, the latent code vectors represent the manifold for cardiac and respiratory motion.
The follow-up 3D approach \mbox{MoCo-SToRM}~\cite{zou2022dynamic} models the image series as a single static 3D template image, which is warped by time-dependent deformation fields.
Instead of generating image frames, the generator network outputs the deformation fields, and the voxels of the template image are considered trainable parameters and are optimized along with the network parameters and code vectors.
This approach was designed for 3D radial lung MRI with a focus on respiratory motion.
More recent DIP approaches for cardiac cine MRI reconstruction by Ahmed et~al., called DEBLUR~\cite{ahmed2022dynamic}, and Hamilton et~al., called LR-DIP~\cite{hamilton2023lowrank}, model the cardiac cine series as a low-rank system, where two separate neural networks are used to generate the 2D spatial and the 1D temporal basis, respectively.

In this work, we present Multi-Dynamic Deep Image Prior (M-DIP)~\cite{vornehm2025motionguided}.
Unlike MoCo-SToRM, M-DIP generates a spatial dictionary comprising multiple elements and synthesizes an intermediate image through a weighted combination of these spatial dictionary elements using time-dependent weights.
M-DIP then models physiological motion as time-dependent deformation fields applied to the intermediate images.
The capability to model both motion and frame-to-frame content changes makes M-DIP suitable for a wide range of applications.
We evaluate M-DIP using data from the MRXCAT cine phantom~\cite{wissmann2014mrxcat}, free-breathing real-time cine, single-shot late gadolinium enhancement (LGE), and first-pass perfusion imaging.

\section{Methods}
\subsection{DIP for dynamic applications}
For an arbitrary variable $(\vec{\cdot})$, we denote its temporal sequence as $(\vec{\cdot})^{(1:T)} \coloneq \left\{(\vec{\cdot})^{(t)}\right\}_{t=1}^{T}$, where $T$ is the total number of frames, and $t$ represents the frame index.
Using this notation, we represent a sequence of image frames, k-space data, and forward operators, as $\vec{x}^{(1:T)}$, $\vec{y}^{(1:T)}$, and $\vec{A}^{(1:T)}$, respectively, such that
\begin{equation}
    \vec{y}^{(t)} = \vec{A}^{(t)}\vec{x}^{(t)} + \vec{\epsilon}^{(t)},
    \label{eq:forward}
\end{equation}
where ${\vec{x}^{(t)} \in \Complex^N}$, ${\vec{y}^{(t)} \in \Complex^M}$, ${\vec{A}^{(t)} \in \Complex^{M \times N}}$, and ${\vec{\epsilon}^{(t)}\in \Complex^M}$ represent the $N$-voxel image, multicoil \mbox{k-space} data, forward operator, and measurement noise, respectively, for the $t^{\text{th}}$ frame.

A straightforward application of DIP for the inverse problem in \eqref{forward} is to solve the following optimization problem:
\begin{equation}
    \Hvec{\xi},\Hvec{z}^{(1:T)} = \argmin_{\vec{\xi}, \vec{z}^{(1:T)}} \sum_{t=1}^T \left\| \vec{A}^{(t)} \Tvec{x}^{(t)} - \vec{y}^{(t)} \right\|_2^2, \label{eq:dip_dynMRI}
\end{equation}
with
\begin{equation*}
    \Tvec{x}^{(t)} = \mathcal{G}_{\vec{\xi}}(\vec{z}^{(t)}),
\end{equation*}
where $\Tvec{x}$ is an estimate of the true image $\vec{x}$, ${\mathcal{G}_{\vec{\xi}} \colon \Real^K \rightarrow \Complex^N}$ is a neural network parameterized by $\vec{\xi}$, and $\vec{z}^{(1:T)}$ represents time-dependent code vectors, with $\vec{z}^{(t)} \in \Real^K$ for $K>0$.
After training, an arbitrary $t^{\text{th}}$ frame can be recovered by $\Hvec{x}^{(t)} = \mathcal{G}_{\Hvec{\xi}}(\Hvec{z}^{(t)})$.
This naive approach, however, does not explicitly model the temporal structure in a dynamic image series $\vec{x}^{(1:T)}$.

\subsection{M-DIP framework}
\begin{figure*}[!t]
    \centering
    \includegraphics[width=\textwidth]{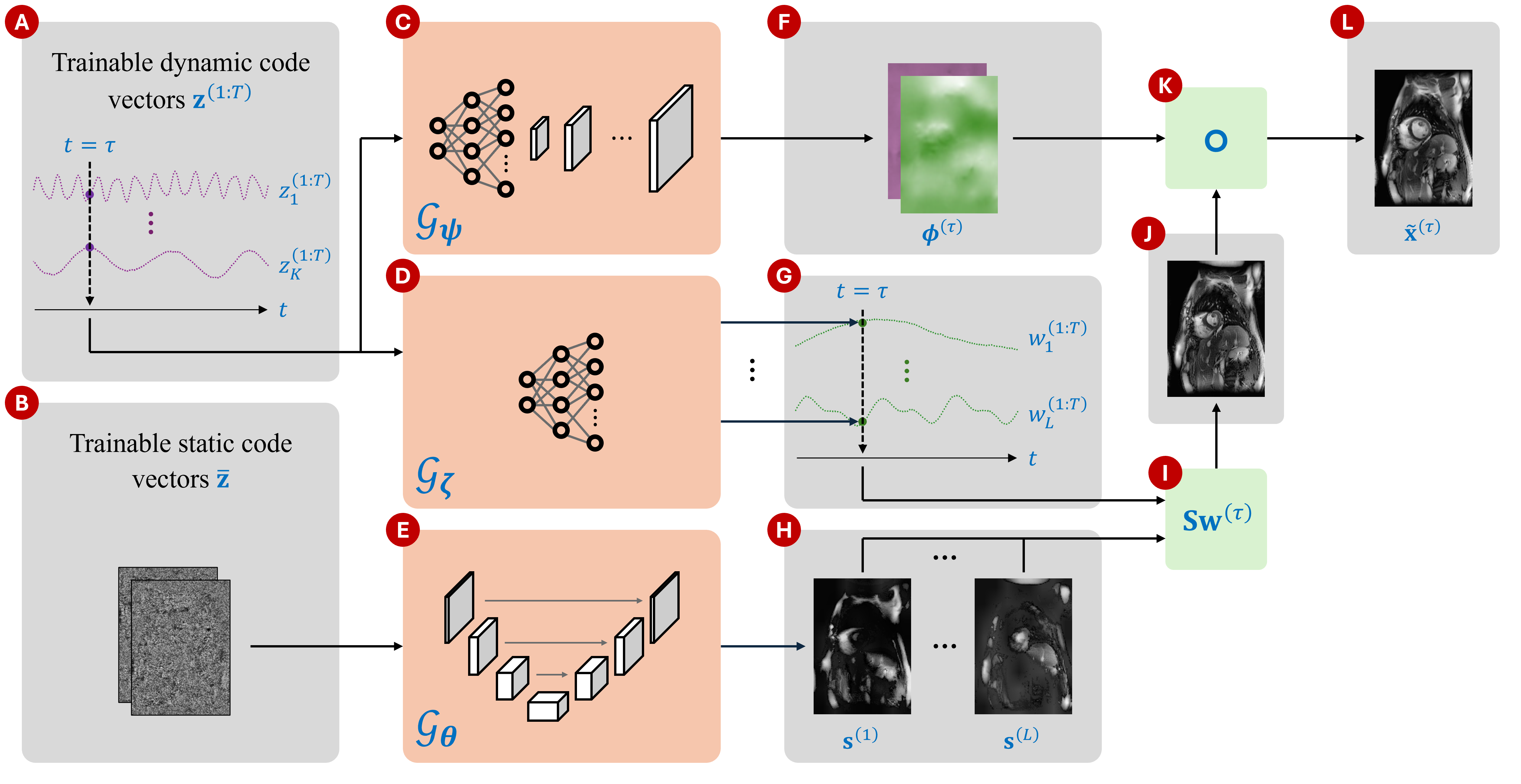}
    \caption{Overview of M-DIP. (A)~Trainable dynamic code vectors of dimensionality~$K$, (B)~trainable static code vectors of dimensionality~$N \times c$, (C)~the network~$\mathcal{G}_{\vec{\psi}}$ that generates a different deformation field for each~$t$, (D)~the network~$\mathcal{G}_{\vec{\zeta}}$ that generates a different set of weights for each~$t$, (E)~the network~$\mathcal{G}_{\vec{\theta}}$ that generates $L$ time-invariant spatial dictionary elements, (F)~x- and y-components of the deformation field generated by $\mathcal{G}_{\vec{\psi}}$ at $t=\tau$, (G)~weights generated by $\mathcal{G}_{\vec{\zeta}}$ at $t=\tau$, (H)~spatial dictionary~$\vec{s}^{(1:L)}$ generated by $\mathcal{G}_{\vec{\theta}}$, (I)~linear combination of the dictionary elements with time-specific weights at $t=\tau$, (J)~intermediate image at $t=\tau$, (K)~warping operation that applies deformation to the intermediate image, and (L)~the output frame at $t=\tau$.}
    \label{fig:overview}
\end{figure*}

Our proposed method draws inspiration from several previously described DIP approaches for dynamic MRI reconstruction.
An overview of the framework is illustrated in \figref{overview}.
In summary, we model a 2D dynamic MRI series as a set of $T$ image frames~$\vec{x}^{(1:T)}$ such that $\vec{x}^{(\tau)}$ at time~$\tau$ is constructed by warping an intermediate image with deformation fields~${\vec{\phi}^{(\tau)} \in \Real^{N \times 2}}$.
These time-dependent deformation fields model in-plane cardiac and respiratory motion.
However, these fields alone cannot effectively model through-plane motion or contrast changes over time.
To model such variations, we synthesize the intermediate image from $L$ spatial dictionary elements ${\vec{s}^{(1:L)} \coloneq \big\{ \vec{s}^{(i)} \big\}_{i=1}^L}$ using weights~${\vec{w}^{(1:T)} \coloneq \big\{ \vec{w}^{(t)} \big\}_{t=1}^T}$, where ${\vec{s}^{(i)} \in \Complex^N}$ and ${\vec{w}^{(t)} \in \Complex^L}$.
The matrix $\vec{S} \in \Complex^{N \times L}$ contains the elements of $\vec{s}^{(1:L)}$ as its columns.
Then, an estimate of the $\tau^{\text{th}}$ frame $\Tvec{x}^{(\tau)}$ is expressed as
\begin{equation}
    \Tvec{x}^{(\tau)} = \underbrace{~\vec{S}\vec{w}^{(\tau)}~}_{\text{\tiny {intermediate image}}} \circ~~ \vec{\phi}^{(\tau)},
    \label{eq:mdip}
\end{equation}
where ``$\circ$'' represents spatial warping~\cite{jaderberg2015spatial}.

The three entities on the right hand side of \eqref{mdip} are generated from three separate neural networks ${\mathcal{G}_{\vec{\theta}} \colon \Real^{N \times c} \rightarrow \Complex^{N \times L}}$, ${\mathcal{G}_{\vec{\zeta}} \colon \Real^K \rightarrow \Complex^L}$, and ${\mathcal{G}_{\vec{\psi}} \colon \Real^K \rightarrow \Real^{N \times 2}}$ such that ${\vec{S} = \mathcal{G}_{\vec{\theta}}(\bvec{z})}$, ${\vec{w}^{(\tau)} = \mathcal{G}_{\vec{\zeta}}(\vec{z}^{(\tau)})}$, and ${\vec{\phi}^{(\tau)} = \mathcal{G}_{\vec{\psi}}(\vec{z}^{(\tau)})}$, where $\vec{\theta}$, $\vec{\zeta}$, and $\vec{\psi}$ are the parameters of the three networks, and ${\bvec{z} \in \Real^{N \times c}}$ represents the static code vector of size~$N$ and with $c$ channels.
The dynamic and static code vectors, $\vec{z}^{(1:T)}$ and $\bvec{z}$, are optimized along with the network parameters $\vec{\theta}$, $\vec{\zeta}$, and $\vec{\psi}$.
In addition, we add regularization to ensure smoothness of the deformation fields $\vec{\phi}^{(1:T)}$.
The optimization problem solved in M-DIP then is:
\begin{equation}
    \Hvec{\theta}, \Hvec{\zeta}, \Hvec{\psi}, \Hat{\bvec{z}}, \Hvec{z}^{(1:T)} = \argmin_{\vec{\theta}, \vec{\zeta}, \vec{\psi}, \bvec{z}, \vec{z}^{(1:T)}}~\sum_{t=1}^{T} \left\| \vec{A}^{(t)} \left[ \vec{S}\vec{w}^{(t)} \circ \vec{\phi}^{(t)} \right] - \vec{y}^{(t)} \right\|_2^2 + \lambda_{\text{s}} \left\| g_{\text{s}}\big( \vec{\phi}^{(1:T)}\big) \right\|_2^2 + \lambda_{\text{f}} \left\| g_{\text{f}}\big( \vec{\phi}^{(1:T)}\big) \right\|_2^2,
    \label{eq:optim}
\end{equation}
with
\begin{equation*}
    \vec{S} = \mathcal{G}_{\vec{\theta}}(\bvec{z}) ,\quad \vec{w}^{(t)} = \mathcal{G}_{\vec{\zeta}}(\vec{z}^{(t)}) ,\quad  \vec{\phi}^{(t)} = \mathcal{G}_{\vec{\psi}}(\vec{z}^{(t)}).
\end{equation*}
The functions $g_{\text{s}}$ and $g_{\text{f}}$ compute the finite differences along the spatial and frame dimensions, respectively, and $\lambda_{\text{s}}$ and $\lambda_{\text{f}}$ are the corresponding regularization weights.

\subsubsection{Spatial dictionary generator}
The spatial dictionary generator network~$\mathcal{G}_{\vec{\theta}}$ follows a U-Net architecture~\cite{ronneberger2015unet}.
It accepts as input the static code vector~$\bvec{z}$ with the number of channels chosen as ${c=2}$.
The network's output consists of $L$ complex-valued images, represented by $2L$ channels, where $L$ is a user-defined parameter dependent on the application.
Note that we consider the output of $\mathcal{G}_{\vec{\theta}}$ not as a basis but as a dictionary, since its elements are not required to be linearly independent and may be overcomplete if $L$ is large.

Each convolutional block in the U-Net comprises two 2D convolutional layers, each followed by a leaky ReLU activation function~\cite{maas2013rectifier} and batch normalization~\cite{ioffe2015batch}.
Downsampling is achieved via average pooling, whereas upsampling is performed using bilinear interpolation.
A detailed network architecture is provided in \figref{architecture_details}.

\subsubsection{Temporal weights generator}
The temporal weights generator network~$\mathcal{G}_{\vec{\zeta}}$ is a multi-layer perceptron (MLP) consisting of seven fully connected layers and leaky ReLU activation functions.
The input size is $K=4$, and the output size is $2L$.
For every code vector~$\vec{z}^{(t)}$, the MLP generates $L$ complex-valued weights~$\vec{w}^{(t)}$.
The network architecture is further detailed in \figref{architecture_details}.

\subsubsection{Deformation field generator}
The deformation field generator network~$\mathcal{G}_{\vec{\psi}}$ accepts the same input as the temporal weights generator.
Initially, an MLP with five fully connected layers expands the latent space from $K$ to $N / 2^{4}$.
The resulting feature vector is reshaped into a 2D matrix and passed through a convolutional neural network with five blocks, separated by four upsampling operations.
Each block consists of three convolutional layers, followed by a leaky ReLU activation function and batch normalization.
Upsampling is performed via nearest-neighbor interpolation.
A final convolutional layer with two output channels generates the deformation fields in the $x$ and $y$ directions.
The detailed network architecture is provided in \figref{architecture_details}.

\subsubsection{Implementation details}
We optimize the three networks jointly for $N_{\text{iter}}$ iterations, with the deformation field generator activated after $N_{\text{def}}$ iterations.
During the first $N_{\text{def}}$ iterations, the warping operation ``$\circ$'' is replaced by the identity mapping, promoting more accurate modeling of content variation not attributable to motion in the intermediate image.
A cosine annealing learning rate schedule~\cite{loshchilov2017sgdr} is used to reduce the learning rate to 0.1\,\% of its initial value over the course of training.
We apply different initial learning rates for the static components (i.\,e., the spatial dictionary generator network~$\mathcal{G}_{\vec{\theta}}$ and static code vector~$\bvec{z}$) and the dynamic components (i.\,e., the temporal weights generator~$\mathcal{G}_{\vec{\zeta}}$, deformation field generator~$\mathcal{G}_{\vec{\psi}}$, and dynamic code vectors~$\vec{z}^{(1:T)}$).
These initial learning rates are denoted as $\eta_{\text{s}}$ for the static components and $\eta_{\text{f}}$ for the dynamic components.

Following Ulyanov et~al.~\cite{ulyanov2020deep}, we initialize the elements of $\bvec{z}$ as independent and identically distributed (i.\,i.\,d.) samples from the uniform distribution $\mathcal{U}\left(0,0.1\right)$.
Additionally, we apply noise regularization~\cite{ulyanov2020deep} by adding i.\,i.\,d. Gaussian noise ${\mathcal{N}\left(0,\sigma_n^2\right)}$ to $\bvec{z}$ at each training iteration~$n$, where $\sigma_n^2$ denotes the noise variance.
This promotes minima that are insensitive to small input perturbations.
The magnitude of the added noise decreases over the course of training, following ${\sigma_n=\sigma_0 (1-0.9\,n/N_{\text{iter}})}$, where $\sigma_0$ is the initial noise level.

In contrast, the dynamic code vectors~$\vec{z}^{(1:T)}$ are initialized to a constant value of zeros.
To optimize memory usage, we process a randomly selected mini-batch of $\min(T, 96)$ consecutive frames in each training iteration.
Prior to network optimization, the data are compressed to 12 virtual coils, and coil sensitivity maps are estimated from time-averaged k-space data using ESPIRiT~\cite{uecker2014espirit}.
All reconstructions were performed on an NVIDIA A100 GPU with 80\,GB of memory.

\subsection{Baseline methods}
We compare our method to LR-DIP~\cite{hamilton2023lowrank}, a state-of-the-art DIP-based approach for reconstructing cardiac cine MRI.
An implementation for LR-DIP was provided by the original authors and adapted for Cartesian data.
We use 5\,\% dropout rate, which was found by the authors to yield the best image quality at 1.5\,T.
The rank~$k_{\text{lr}}$ is selected based on the application and chosen to be considerably lower than the number of image frames $T$.
The default depth of the spatial and temporal U-Net in \mbox{LR-DIP}, i.\,e., the number of downsampling and upsampling operations, is five.
We use this value in our experiments unless the dimensionality of the data necessitates a lower value.

Additionally, we compare our method to reconstructions obtained using the low-rank plus sparse (L+S) method~\cite{otazo2015lowrank}.
Reconstructions of in-vivo cine data are furthermore compared to a supervised deep learning reconstruction for cardiac cine MRI, called CineVN~\cite{vornehm2025cinevn}.

\subsection{Experiments and data}
To evaluate M-DIP, we performed experiments with phantom and clinical patient data.
For the human subject data, approval was granted by the Institutional Review Board (IRB) at The Ohio State University (2019H0076).
Informed consent was obtained from all individual participants included in this work.

\subsubsection{Phantoms}
Seven datasets from real-time cardiac cine MRXCAT phantoms with varying anatomies were simulated~\cite{wissmann2014mrxcat}, with an isotropic in-plane resolution of 2\,mm and a slice thickness of 8\,mm.
Each simulated cine was 9\,s long, containing two breathing cycles and 10 cardiac cycles of random durations between 0.8\,s and 1.0\,s.
Motion was simulated in 3D phantoms before extraction of 2D slices to ensure presence of both in-plane and through-plane motion.
At a temporal resolution of 30\,ms, each cine series consisted of 300~image frames.
Complex multicoil k-space data for 12 coils were simulated, and random complex-valued Gaussian noise was added to achieve a signal-to-noise ratio of 10\,dB in the individual coil images.
The simulated k-space data were then retrospectively undersampled using a variable density golden ratio offset (GRO) Cartesian sampling pattern~\cite{joshi2022technical} with an acceleration factor of eight.

Reconstructions were performed using L+S, \mbox{LR-DIP}, and the proposed \mbox{M-DIP}, and were compared to the noise-free ground-truth simulation in terms of peak signal-to-noise ratio (PSNR), normalized root mean square error (NRMSE), and structural similarity index (SSIM).
We use a dictionary size of ${L=16}$ and train for ${N_{\text{iter}}=10\,000}$ iterations.
Further hyperparameters are provided in \tabref{hyperparameters}.

\subsubsection{Real-time cine}
Free-breathing real-time cine data were collected from 27 unique clinical patients on a 1.5\,T scanner (MAGNETOM Sola, Siemens Healthineers, Forchheim, Germany) using a balanced steady-state free precession sequence.
The acquisitions were prospectively undersampled using a GRO sampling pattern with an acceleration rate between seven and ten.
Between 196 and 212 phases were acquired in each cine with a temporal resolution of \mbox{47--51\,ms}.
Each cine was 10\,s long, typically covering two to four breathing cycles and 10 to 15 cardiac cycles.
In-plane resolution varied between 2.04\,mm and 2.84\,mm and the slice thickness between 6\,mm and 8\,mm.
A flip angle of 70\textdegree\ was used in all protocols.
Twenty-six slices were acquired in the short-axis view and one in the long-axis view.

Reconstructions were performed using CineVN, L+S, \mbox{LR-DIP}, and the proposed \mbox{M-DIP}.
We use a dictionary size of ${L=16}$ and train for ${N_{\text{iter}}=10\,000}$ iterations.
Further hyperparameters are provided in \tabref{hyperparameters}.
Two readers with more than ten years of experience in CMR blindly scored all reconstructions in terms of ``image sharpness'' and ``perceived noise and artifacts'' on five-point Likert scales (1=Nondiagnostic, 2=Poor, 3=Fair, 4=Good, 5=Excellent).

An additional experiment was performed to demonstrate the role of deformation fields in modeling motion in M-DIP.
The deformation process was omitted by replacing the warping operation ``$\circ$'' with the identity mapping.
The number of parameters in the temporal weights generator~$\mathcal{G}_{\vec{\zeta}}$ and spatial dictionary generator~$\mathcal{G}_{\vec{\theta}}$ was increased to match the model capacity of the previous experiments, and the dictionary size was doubled to ${L=32}$.

\subsubsection{Single-shot LGE}
Thirty-three free-breathing single-shot LGE image series were collected from 20 unique clinical patients on the same 1.5\,T scanner.
In this patient group, performing breath-held segmented LGE was not feasible due to arrhythmias or patients' inability to hold breath.
A phase-sensitive inversion recovery (PSIR) sequence~\cite{kellman2002phasesensitive} with inversion times between 270\,ms and 410\,ms and a flip angle of 40\textdegree\ was used.
In-plane resolution varied between 1.37\,mm and 1.88\,mm, with a slice thickness of 8\,mm.
A GRO sampling pattern was used for prospective undersampling with an acceleration rate between four and six.
Each series consisted of 32 T1-weighted repetitions, each representing a single-shot acquisition that was prospectively triggered using a proprietary ECG signal.
The temporal footprint of each repetition was 120--135\,ms.
Fifteen slices were acquired in the short-axis view and 18 in the long-axis view.

Reconstructions were performed using L+S, \mbox{LR-DIP}, and the proposed \mbox{M-DIP}.
Note that we use a lower dictionary size of $L=8$ compared to the cine data due to the lower number of image frames in the LGE series.
Similarly, we reduce the rank $k_{\text{lr}}$ in LR-DIP to $12$ and the depth of the temporal basis network in LR-DIP to four for the same reason.
Furthermore, we do not use regularization along the temporal dimension of the deformation fields in M-DIP ($\lambda_{\text{f}}=0$) because consecutive frames are not expected to be similar in their breathing motion state.
Detailed hyperparameters are provided in \tabref{hyperparameters}.
Reconstructions were scored in terms of ``clarity of pertinent myocardial features'' on a five-point Likert scale (1=Nondiagnostic, 2=Poor, 3=Fair, 4=Good, 5=Excellent).

\subsubsection{First-pass perfusion}
The perfusion data comprised 23 image series collected from 12 unique clinical patients on the same 1.5\,T scanner as the previous datasets.
A flip angle of 12\textdegree\ was used with $T_{\text{E}}$ and $T_{\text{R}}$ of 1.1--1.5\,ms and 2.5--3.3\,ms, respectively.
In-plane resolution varied between 2.25\,mm and 2.81\,mm, with a slice thickness of 8--10\,mm.
The acceleration rate ranged between 5.0 and 6.3 using the GRO sampling pattern, and each series contained 60 ECG-triggered repetitions.
The temporal footprint of each repetition was 74--98\,ms.
Nineteen slices were acquired in the short-axis view and four in the long-axis view.

Reconstructions were performed using L+S, \mbox{LR-DIP}, and the proposed \mbox{M-DIP}.
We use a dictionary size of ${L=24}$ and train for ${N_{\text{iter}}=8\,000}$ iterations, where the deformation field generator is activated after ${N_{\text{def}}=1\,000}$ iterations.
Note that the number of dictionary elements~$L$ relative to the number of image frames is higher for perfusion than for cine and LGE data.
This accounts for the expected larger content variation in the data due to the contrast change over time.
Further hyperparameters are provided in \tabref{hyperparameters}.
Reconstructions were scored in terms of ``clarity of contrast dynamics and myocardial defects'' on a five-point Likert scale (1=Nondiagnostic, 2=Poor, 3=Fair, 4=Good, 5=Excellent).

An additional experiment was performed to demonstrate the role of the dictionary in modeling of contrast dynamics in M-DIP.
The dictionary size was set to ${L=1}$, the temporal weights generator~$\mathcal{G}_{\vec{\zeta}}$ was omitted, and the weights set to a scalar ${w^{(\tau)}=1}$ for every time point~$\tau$.
The number of parameters in the deformation field generator~$\mathcal{G}_{\vec{\psi}}$ was increased to match the model capacity of the previous experiments.

\subsubsection{Motion compensation}
The M-DIP framework enables reconstruction of motion-compensated image series, provided that in-plane motion and other variations are modeled separately by the deformation fields and the spatial dictionary, respectively.
This is implemented by using deformation fields of one selected frame $\vec{\phi}^{(\tau)}$ to deform all $T$ frames during inference.
For demonstration, reconstructions of the in-vivo data were repeated using this approach, with $\tau=0$ chosen as the reference frame.

\section{Results}
\subsection{Phantom study}
\begin{figure}[!t]
    \centering
    \includegraphics[width=13cm]{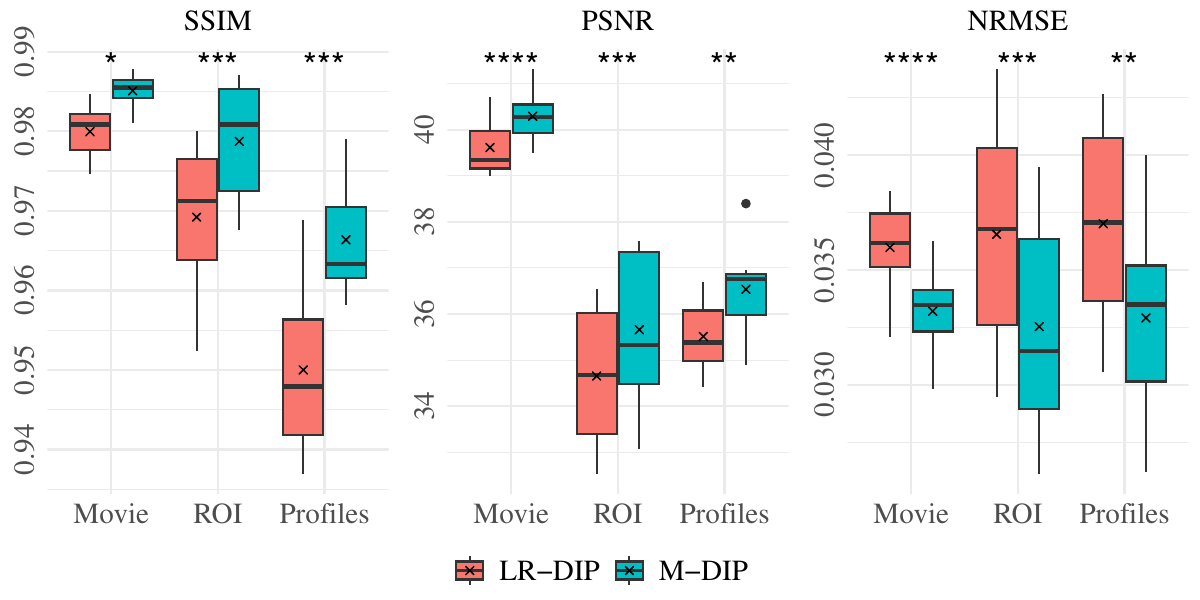}
    \caption{M-DIP and LR-DIP results of the phantom study with results of Student's t-tests. $\texttt{*}$ indicates ${p\leq0.05}$, $\texttt{**}$ indicates ${p\leq0.01}$, $\texttt{***}$ indicates ${p\leq0.001}$, and $\texttt{****}$ indicates ${p\leq0.0001}$.}
    \label{fig:mrxcat_results_boxplots}
\end{figure}

\begin{figure*}[!t]
    \centering
    \begin{subfigure}[t]{0.23\textwidth}
        \centering
        \includegraphics[width=\textwidth]{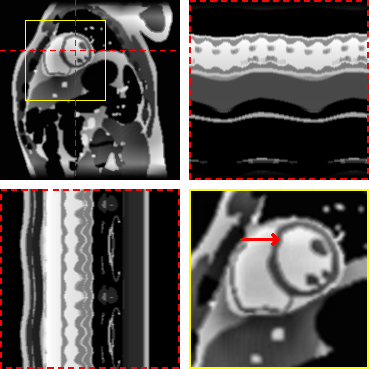}
        \caption{Ground truth}
    \end{subfigure}
    \hspace{2mm}
    \begin{subfigure}[t]{0.23\textwidth}
        \centering
        \includegraphics[width=\textwidth]{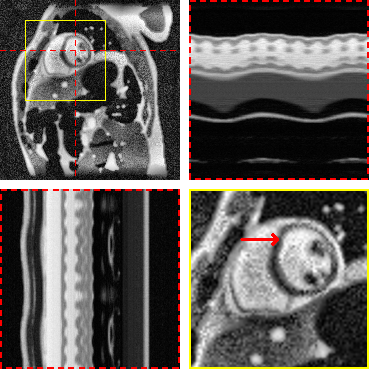}
        \caption{L+S}
    \end{subfigure}
    \hspace{2mm}
    \begin{subfigure}[t]{0.23\textwidth}
        \centering
        \includegraphics[width=\textwidth]{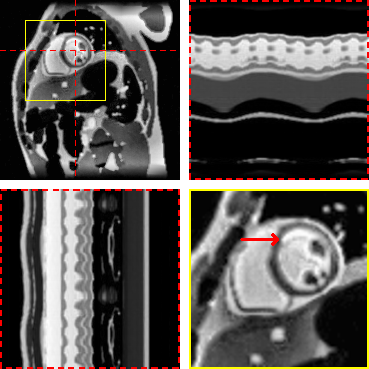}
        \caption{LR-DIP}
    \end{subfigure}
    \hspace{2mm}
    \begin{subfigure}[t]{0.23\textwidth}
        \centering
        \includegraphics[width=\textwidth]{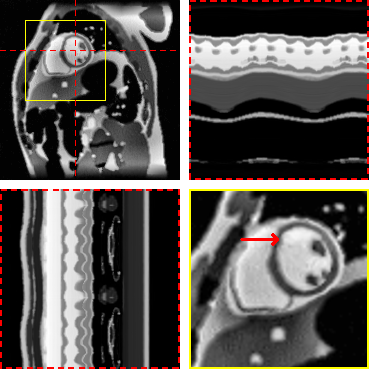}
        \caption{M-DIP}
    \end{subfigure}
    \caption{Exemplary MRXCAT phantom reconstructions. Each sub-figure illustrates an end-diastolic frame, temporal profiles, and a close-up of the heart. Red arrows highlight an artifact present in LR-DIP that is suppressed in M-DIP.}
    \label{fig:mrxcat_example}
\end{figure*}

The results of the phantom study are provided in \tabref{mrxcat_results}.
SSIM, PSNR, and NRMSE were computed on the full cine movies, on a region of interest (ROI) around the heart, and on temporal profiles through the center of the heart.
M-DIP achieved the best scores in all metrics.
The results of M-DIP and LR-DIP are further illustrated in \figref{mrxcat_results_boxplots}, with Student's t-tests conducted to assess statistical significance.
All reported $p$-values are significant ($\alpha=0.05$) after correction for multiple comparisons using the Holm-Bonferroni method.
Reconstruction times were approximately 40\,min for M-DIP and 70\,min for LR-DIP.
Exemplary reconstructions of one phantom are shown in \figref{mrxcat_example}.

\begin{table}[!t]
    \centering
    \caption{Results of the phantom study, with the best value in each column highlighted in bold.}
    \label{tab:mrxcat_results}
    \footnotesize
    \begin{tabularx}{8cm}{
        @{\hspace{5pt}}
        cr|
        @{\extracolsep\fill}
        S[table-format=1.3] S[table-format=2.1,table-space-text-post={\,dB}] S[table-format=1.4]
        @{\extracolsep\fill}
        @{\hspace{5pt}}
    }\toprule
        &        &   {SSIM} &   {PSNR}    &   {NRMSE} \\\midrule
        \multirow{3}{*}{\rotatebox[origin=c]{90}{Movie}}
        &    L+S &    0.700 &    30.0\,dB &    0.1082 \\
        & LR-DIP &    0.980 &    39.6\,dB &    0.0360 \\
        &  M-DIP & \bf0.985 & \bf40.3\,dB & \bf0.0332 \\\midrule
        \multirow{3}{*}{\rotatebox[origin=c]{90}{ROI}}
        &    L+S &    0.824 &    26.1\,dB &    0.1015 \\
        & LR-DIP &    0.969 &    34.7\,dB &    0.0366 \\
        &  M-DIP & \bf0.979 & \bf35.7\,dB & \bf0.0325 \\\midrule
        \multirow{3}{*}{\rotatebox[origin=c]{90}{Profiles}}
        &    L+S &    0.652 &    26.7\,dB &    0.1055 \\
        & LR-DIP &    0.950 &    35.5\,dB &    0.0370 \\
        &  M-DIP & \bf0.966 & \bf36.5\,dB & \bf0.0329 \\\bottomrule
    \end{tabularx}
\end{table}

\subsection{In-vivo study}
\begin{figure}[!t]
    \centering
    \includegraphics[width=13cm]{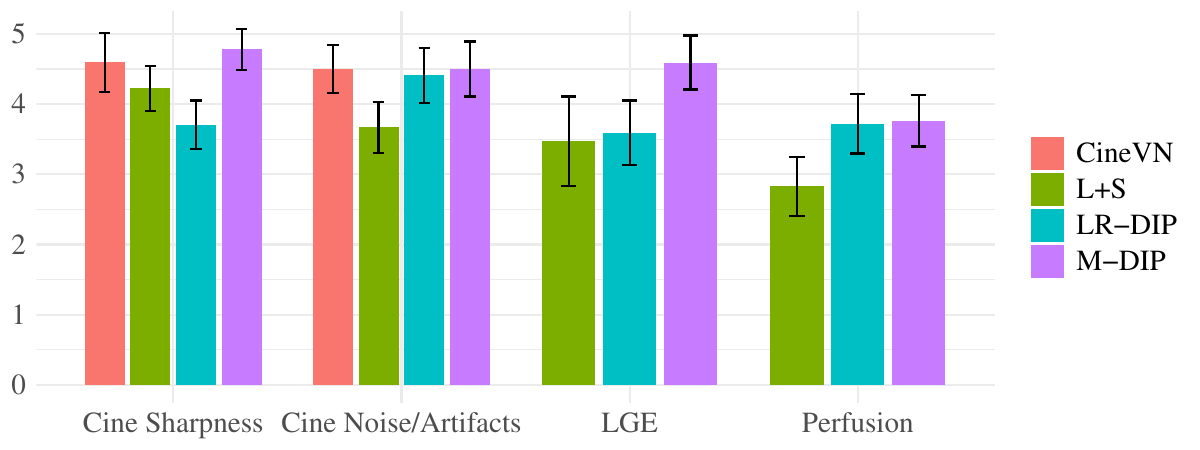}
    \caption{Bar plots with the scoring results for the in-vivo studies.}
    \label{fig:scoring}
\end{figure}

\begin{figure*}[!t]
    \centering
    \begin{subfigure}[t]{0.23\textwidth}
        \centering
        \includegraphics[width=\textwidth]{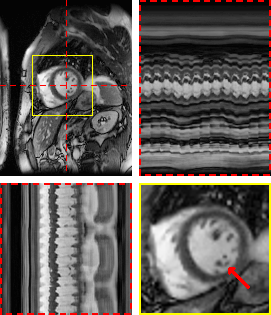}
        \caption{CineVN (supervised)}
    \end{subfigure}
    \hspace{2mm}
    \begin{subfigure}[t]{0.23\textwidth}
        \centering
        \includegraphics[width=\textwidth]{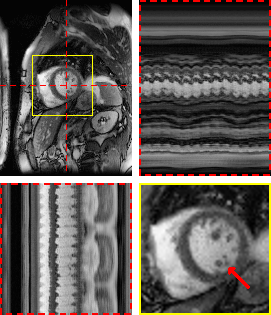}
        \caption{L+S}
    \end{subfigure}
    \hspace{2mm}
    \begin{subfigure}[t]{0.23\textwidth}
        \centering
        \includegraphics[width=\textwidth]{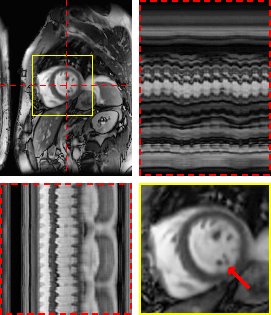}
        \caption{LR-DIP}
    \end{subfigure}
    \hspace{2mm}
    \begin{subfigure}[t]{0.23\textwidth}
        \centering
        \includegraphics[width=\textwidth]{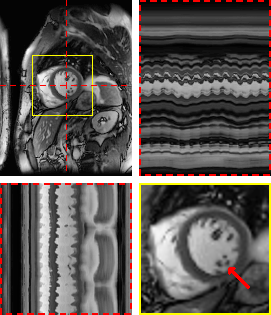}
        \caption{M-DIP}
    \end{subfigure}
    \caption{Exemplary in-vivo real-time cine reconstructions. Each sub-figure illustrates an end-diastolic frame, temporal profiles, and a close-up of the heart. Red arrows show an area where differences in image sharpness are particularly apparent.}
    \label{fig:cine_example}
\end{figure*}

\begin{figure*}[!t]
    \centering
    \begin{subfigure}[t]{0.23\textwidth}
        \centering
        \includegraphics[width=\textwidth]{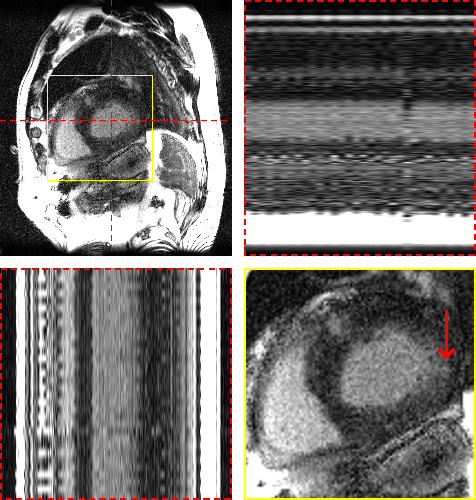}
        \caption{L+S}
    \end{subfigure}
    \hspace{2mm}
    \begin{subfigure}[t]{0.23\textwidth}
        \centering
        \includegraphics[width=\textwidth]{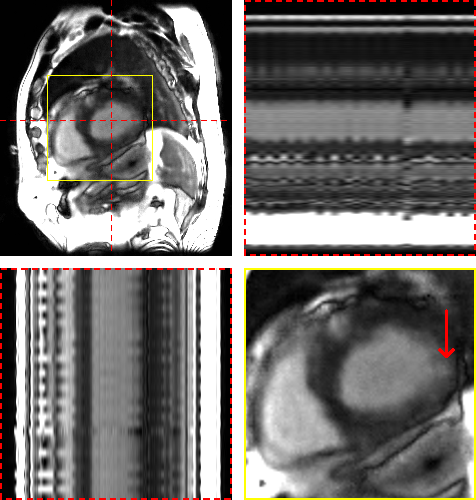}
        \caption{LR-DIP}
    \end{subfigure}
    \hspace{2mm}
    \begin{subfigure}[t]{0.23\textwidth}
        \centering
        \includegraphics[width=\textwidth]{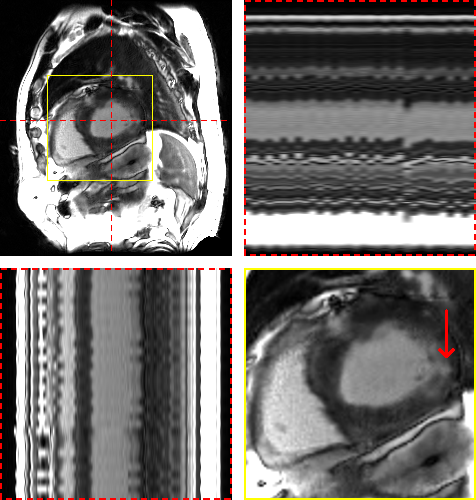}
        \caption{M-DIP}
    \end{subfigure}
    \caption{Exemplary in-vivo free-breathing single-shot LGE reconstructions. Each sub-figure illustrates one frame, temporal profiles, and a close-up of the heart. Red arrows show an enhanced area that is better visible in M-DIP compared to LR-DIP.}
    \label{fig:lge_example}
\end{figure*}

\begin{figure*}[!t]
    \centering
    \begin{subfigure}[t]{0.23\textwidth}
        \centering
        \includegraphics[width=\textwidth]{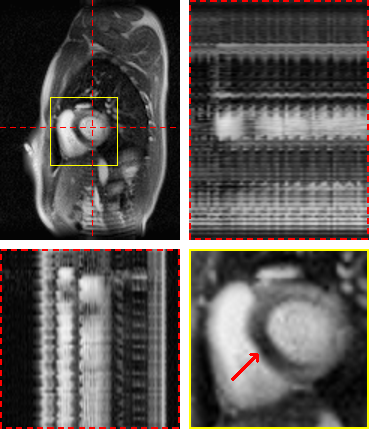}
        \caption{L+S}
    \end{subfigure}
    \hspace{2mm}
    \begin{subfigure}[t]{0.23\textwidth}
        \centering
        \includegraphics[width=\textwidth]{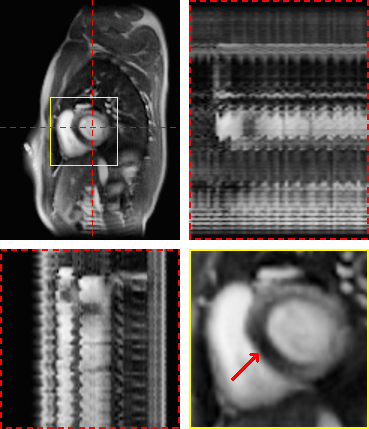}
        \caption{LR-DIP}
    \end{subfigure}
    \hspace{2mm}
    \begin{subfigure}[t]{0.23\textwidth}
        \centering
        \includegraphics[width=\textwidth]{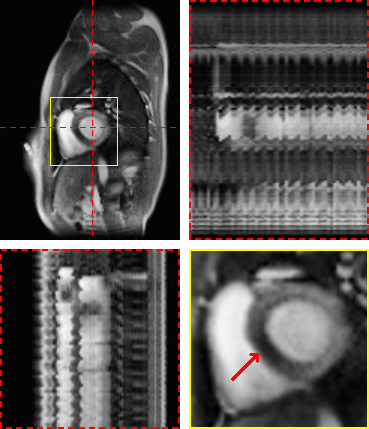}
        \caption{M-DIP}
    \end{subfigure}
    \caption{Exemplary in-vivo first-pass perfusion reconstructions. Each sub-figure illustrates one frame, temporal profiles, and a close-up of the heart. Red arrows show a septal perfusion defect clearly visible in all three reconstructions.}
    \label{fig:perfusion_example}
\end{figure*}

Scoring results are summarized in \tabref{scoring} and \figref{scoring}.
Figures~\ref{fig:cine_example} to \ref{fig:perfusion_example} present results for exemplary real-time cine, free-breathing single-shot LGE, and first-pass perfusion datasets, respectively.

\begin{table}[!t]
    \centering
    \caption{Mean scoring results for the in-vivo studies. $p$-values of Student's t-tests with respect to M-DIP are given, where \textsuperscript{(ns)} indicates ${p>0.05}$, * indicates ${p\leq0.05}$, ** indicates ${p\leq0.01}$, and *** indicates ${p\leq0.001}$. The best value in each column is written in boldface.}
    \label{tab:scoring}
    \footnotesize
    \begin{tabularx}{10cm}{
        @{\hspace{5pt}}
        @{\extracolsep\fill}
        r|
        S[table-format=1.2,table-space-text-post={\nonsig}] S[table-format=1.2,table-space-text-post={\nonsig}]|
        S[table-format=1.2,table-space-text-post={\nonsig}]|
        S[table-format=1.2,table-space-text-post={\nonsig}]
        @{\extracolsep\fill}
        @{\hspace{5pt}}
    }\toprule
               & \multicolumn{2}{c|}{Real-time cine} &     {LGE}     &    {Perfusion}    \\
               & {Sharpness} &   {Noise/Artifacts}   &               &                   \\\midrule
        CineVN &    4.59\p   &      \bf4.50\nonsig   &     {---}     &       {---}       \\
           L+S &    4.22\ppp &         3.67\ppp      &     3.47\ppp  &       2.83\ppp    \\
        LR-DIP &    3.70\ppp &         4.41\nonsig   &     3.59\ppp  &       3.72\nonsig \\
         M-DIP & \bf4.78     &      \bf4.50          &  \bf4.59      &    \bf3.76        \\\bottomrule
    \end{tabularx}
\end{table}

In all three applications, M-DIP consistently achieved the highest scores across all evaluated criteria.
Notably, the sharpness of real-time cine images was rated significantly higher for M-DIP compared to any other method.
For noise/artifacts in real-time cine images, M-DIP and CineVN received identical average scores, whereas L+S scored significantly lower in both sharpness and noise/artifacts categories.
For free-breathing single-shot LGE, M-DIP was rated significantly higher than both LR-DIP and L+S.
In the perfusion images, M-DIP and LR-DIP received comparable scores, with M-DIP scoring slightly but not significantly higher, whereas L+S was rated significantly lower.

Reconstruction times varied between methods: Cine reconstruction required approximately 30\,min with \mbox{M-DIP} and 45\,min with LR-DIP; LGE reconstruction took approximately 15\,min with M-DIP and 10\,min with LR-DIP; and perfusion reconstruction required approximately 25\,min with M-DIP and 20\,min with LR-DIP.

\begin{figure*}[!t]
    \centering
    \begin{subfigure}[t]{0.23\textwidth}
        \centering
        \includegraphics[width=\textwidth]{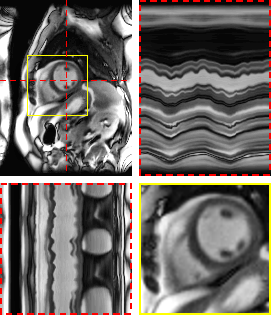}
        \caption{With deformation}
        \label{fig:def_on}
    \end{subfigure}
    \hspace{1mm}
    \begin{subfigure}[t]{0.23\textwidth}
        \centering
        \includegraphics[width=\textwidth]{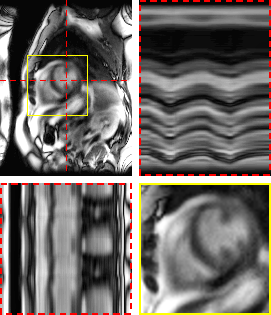}
        \caption{Without deformation}
        \label{fig:def_off}
    \end{subfigure}
    \hspace{4mm}
    \begin{subfigure}[t]{0.23\textwidth}
        \centering
        \includegraphics[width=\textwidth]{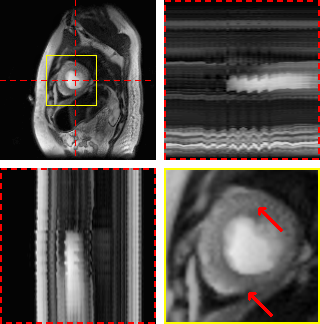}
        \caption{Dictionary size $L=24$}
        \label{fig:dict_size_24}
    \end{subfigure}
    \hspace{1mm}
    \begin{subfigure}[t]{0.23\textwidth}
        \centering
        \includegraphics[width=\textwidth]{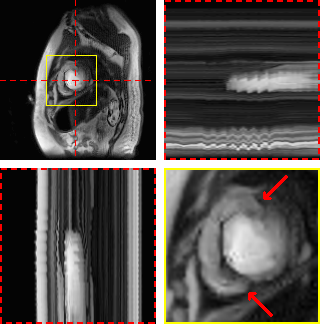}
        \caption{Dictionary size $L=1$}
        \label{fig:dict_size_1}
    \end{subfigure}
    \caption{M-DIP reconstructions of a real-time cine series with the deformation field generator $\mathcal{G}_{\vec{\psi}}$ (a) activated and (b) deactivated, and M-DIP reconstructions of a perfusion dataset with spatial dictionaries of (c)~size $L=24$ and (d)~size $L=1$. The red arrows point to the artifact observed in the perfusion case when $L=1$.}
    \label{fig:def_onoff_dict_size}
\end{figure*}

Figures~\ref{fig:def_on} and \ref{fig:def_off} show the effect of omitting the deformation field generator $\mathcal{G}_{\vec{\psi}}$ on one of the real-time cine series.
Figures~\ref{fig:dict_size_24} and \ref{fig:dict_size_1} demonstrate the effect of learning only one image instead of a dictionary of images when modeling content or contrast variations, as seen in perfusion imaging.

Motion-compensated reconstructions of the in-vivo examples are provided in \figref{moco}.

\section{Discussion}
In this work, we propose a DIP-inspired method called M-DIP, and evaluate it using both simulated and patient data for real-time cine, as well as patient data for single-shot LGE and first-pass perfusion imaging.
M-DIP is an instance-specific approach that does not require training data.
Unlike other DIP methods, M-DIP can model both motion and content variations, making it suitable for a wide range of dynamic applications.

In the first study, we simulated multicoil undersampled k-space data using free-breathing real-time cardiac cine MRXCAT phantoms.
M-DIP outperformed both L+S and the recently proposed LR-DIP method in terms of PSNR, NRMSE, and SSIM.
The performance gap between M-DIP and L+S was substantial, with L+S producing images with high levels of noise.
Although the performance advantage of M-DIP over LR-DIP was more moderate, visual inspection revealed that \mbox{LR-DIP} consistently introduced motion blurring around the myocardium.
In contrast, M-DIP captured cardiac motion more accurately, as shown in \figref{mrxcat_example}.

In the second study, we reconstructed prospectively undersampled free-breathing real-time cine data in clinical patients using M-DIP, LR-DIP, L+S, and a recently proposed supervised learning method (CineVN).
As no ground truth was available, a reader study was performed by two experienced readers.

In terms of noise and artifacts, M-DIP and CineVN performed the best, with LR-DIP receiving insignificantly lower scores.
L+S received the lowest scores, exhibiting high levels of noise in the reconstructions.
In terms of image sharpness, M-DIP reconstructions received significantly higher scores than all other methods, whereas LR-DIP received the lowest scores.
Visual inspection again revealed blurring in the LR-DIP reconstructions, particularly around the myocardium, indicating a limitation of the low-rank model when dealing with complex localized motion in a large field of view, as seen in \figref{cine_example}.
In contrast, the M-DIP framework generates a high-resolution intermediate image for each frame, which is then deformed to the respective motion state.
The high scores for noise and artifacts in M-DIP and LR-DIP further highlight the implicit capability of DIP-based methods to generate natural-looking images.\cite{ulyanov2018deep}

Notably, M-DIP significantly outperformed CineVN in terms of image sharpness, despite not benefiting from training data.
Since acquiring fully-sampled training data for real-time cine is infeasible, CineVN was exclusively trained on breath-held segmented acquisitions, which may have limited its performance on free-breathing real-time data.
In contrast, M-DIP is unsupervised and directly learns the physiological motion and content variations from the undersampled data itself.

We further applied M-DIP to free-breathing single-shot LGE and cardiac perfusion data, demonstrating the method's suitability for dynamic applications beyond cine imaging.
The clarity of myocardial features in the LGE data was rated highest for M-DIP, with a substantial performance gap over LR-DIP and L+S.
Similar to real-time cine, LR-DIP reconstructions were blurred, whereas L+S reconstructions exhibited high levels of noise.
For first-pass perfusion, the reconstructions were rated very similarly for M-DIP and LR-DIP, showing no advantage of our method over previous DIP-based methods.
We attribute this to the nature of the perfusion data, where the temporal variations are dominated by contrast changes instead of motion.
Therefore, the benefit of including deformation fields in M-DIP does not result in a significant performance gain over LR-DIP.

The versatility of the proposed M-DIP is rooted in combining a dictionary approach and deformation fields to model generic dynamic MRI series, as demonstrated by the results in \figref{def_onoff_dict_size}.
When trained without the deformation field generator, the cardiac motion in the real-time cine could not be fully restored due to the limited capability of the spatial dictionary to model complex deformations composed of both respiratory and cardiac motion, despite the increased dictionary size.
Conversely, the perfusion example shows that with only one dictionary element, i.\,e., $L=1$, deformation fields alone are ineffective in modeling the contrast variations.

Our framework generally allows motion-compensated reconstruction by applying the same deformation fields to the intermediate images.
However, this requires that all in-plane motion is strictly modeled by the deformation fields, while through-plane motion and contrast changes are strictly modeled by the spatial dictionary.
The results in \figref{moco} show that this strategy performed well for cine and LGE data but only partially compensated for motion in the perfusion example.
The large number of dictionary elements used for perfusion reconstruction allowed the dictionary to capture some motion-related variation.
To achieve more complete disentanglement of motion from other dynamics, additional constraints or architectural refinements will be required within M-DIP.

One limitation of this work, and of DIP-based image reconstruction methods in general, is the long reconstruction time of several tens of minutes for one image series on a single GPU.
Reconstruction time in \mbox{M-DIP} is primarily influenced by the number of iterations $N_{\text{iter}}$, dictionary size $L$, pixels $N$, and frames $T$.
Additionally, optimizing M-DIP with its three subnetworks requires careful selection of hyperparameters.
Future efforts will focus on reducing the computational demand of M-DIP and extending it to other CMR applications, including mapping.
Furthermore, we will explore other variants of regularization and diffeomorphic warping based on stationary velocity estimation, integrated via scaling and squaring~\cite{dalca2019unsupervised}.

\section{Conclusions}
We have proposed, implemented, and evaluated M-DIP, an unsupervised method for dynamic image reconstruction from highly undersampled data.
Results from simulated cine data demonstrate that M-DIP outperforms competing methods in terms of image quality.
In evaluations on clinical data, M-DIP matches or exceeds the performance of other unsupervised methods across cine, LGE, and perfusion data, and is comparable to a state-of-the-art supervised approach in real-time cine imaging.
By accurately modeling both motion and content variations, M-DIP proves applicable to a wide range of dynamic MRI applications.
Its ability to achieve performance on par with supervised methods without requiring fully sampled training data makes it a promising tool especially for real-time imaging.

\section*{Conflict of interest statement}
MV is an employee of Siemens Healthineers AG.
FK receives research funding from Siemens Healthineers AG, receives patent royalties for AI for MR image reconstruction from Siemens Healthineers AG, holds stock options from Subtle Medical Inc. and serves as scientific advisor to Imaginostics Inc.

\section*{Data availability statement}
Source code is available on GitHub: \url{https://github.com/marcvornehm/M-DIP}

\section*{Acknowledgment}
The authors gratefully acknowledge Jesse Hamilton for providing source code for LR-DIP.

\newpage
\printbibliography

\newpage
\section*{Supporting information}

\setcounter{figure}{0}
\renewcommand{\thefigure}{S\arabic{figure}}
\setcounter{table}{0}
\renewcommand{\thetable}{S\arabic{table}}

\begin{table}[ht]
    \centering
    \caption{Reconstruction hyperparameters.
    M-DIP parameters: Number of dictionary elements ($L$), regularization weights for spatial and temporal smoothness of the deformation fields ($\lambda_{\text{s}}$ and $\lambda_{\text{f}}$, respectively), initial noise regularization level ($\sigma_0$), initial learning rates for static and dynamic model components ($\eta_{\text{s}}$ and $\eta_{\text{f}}$, respectively), number of training iterations ($N_{\text{iter}}$), and number of iterations after which deformation fields are generated ($N_{\text{def}}$).
    LR-DIP parameters: Rank of the low-rank system ($k_{\text{lr}}$) and depth of the spatial and temporal basis U-Nets ($d_{\text{S}}$ and $d_{\text{T}}$, respectively).
    L+S parameters: Regularization parameters for the low-rank ($\lambda_{\text{L}}$) and sparse ($\lambda_{\text{S}}$) component, respectively.}
    \label{tab:hyperparameters}
    \begin{tabularx}{\columnwidth}{
        @{\hspace{5pt}}
        @{\extracolsep\fill}
        l|
        S[table-format=2.0] S S S S[table-format=1e3,exponent-product={\cdot}] S[table-format=1e3,exponent-product={\cdot}] S[table-format=5.0] S[table-format=4.0]|
        S[table-format=2.0] S[table-format=1.0] S[table-format=1.0]|
        S S
        @{\extracolsep\fill}
        @{\hspace{5pt}}
    }\toprule
                    & \multicolumn{8}{c|}{M-DIP} & \multicolumn{3}{c|}{LR-DIP} & \multicolumn{2}{c}{L+S}                               \\
         Dataset    & {$L$} & {$\lambda_{\text{s}}$}
                                    & {$\lambda_{\text{f}}$}
                                            & {$\sigma_0$}
                                                    & {$\eta_{\text{s}}$}
                                                            & {$\eta_{\text{f}}$}
                                                                    & {$N_{\text{iter}}$}
                                                                            & {$N_{\text{def}}$}
                                                                                    & {$k_{\text{lr}}$}
                                                                                            & {$d_{\text{S}}$}
                                                                                                    & {$d_{\text{T}}$}
                                                                                                            & {$\lambda_{\text{L}}$} 
                                                                                                                    & {$\lambda_{\text{S}}$} \\\midrule
         Phantom    & 16    & 0.02  & 0.02  & 0.01  & 1e-3  & 1e-3  & 10000 & 0     & 64    & 5     & 5     & 0.5   & 0.05            \\
         Cine       & 16    & 0.1   & 0.05  & 0.05  & 1e-3  & 1e-3  & 10000 & 0     & 64    & 5     & 5     & 0.5   & 0.05            \\
         LGE        & 8     & 0.1   & 0     & 0.1   & 5e-4  & 1e-3  & 10000 & 0     & 12    & 5     & 4     & 0.01  & 0.1             \\
         Perfusion  & 24    & 0.2   & 0.02  & 0.05  & 3e-4  & 6e-3  & 8000  & 1000  & 24    & 5     & 5     & 0.01  & 0.5             \\\bottomrule
    \end{tabularx}
\end{table}

\begin{figure}[ht]
    \centering
    \begin{subfigure}[t]{\textwidth}
        \centering
        \includegraphics[width=0.85\textwidth]{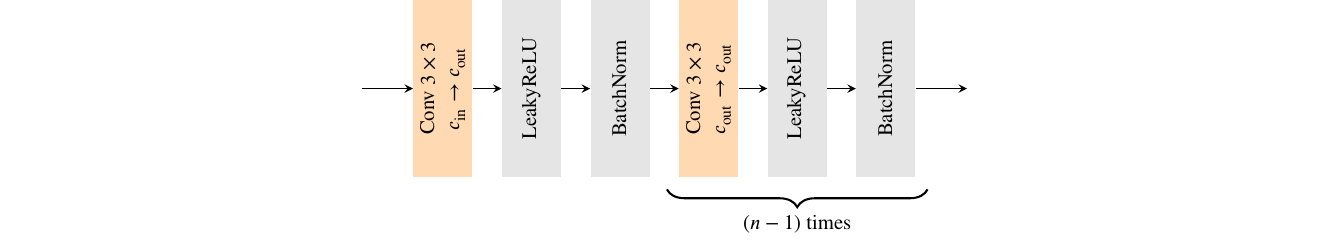}
        \caption{Convolutional block $\operatorname{ConvBlock}(n,c_{\text{in}},c_{\text{out}})$. $n$, $c_{\text{in}}$, and $c_{\text{out}}$ are the number of convolutional layers, input channels, and output channels, respectively.}
    \end{subfigure}
    \vskip2\baselineskip
    \begin{subfigure}[t]{\textwidth}
        \centering
        \includegraphics[width=0.85\textwidth]{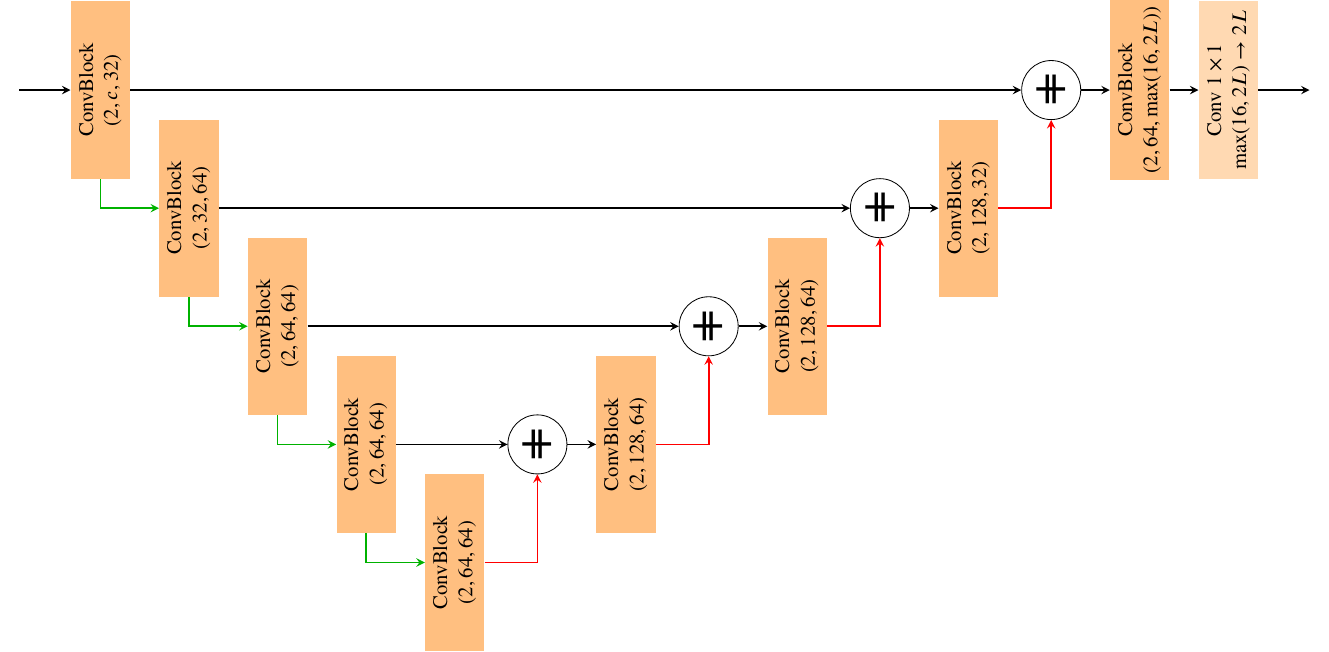}
        \caption{Spatial dictionary generator $\mathcal{G}_{\vec{\theta}}$. Green arrows denote average pooling with kernel size $2 \times 2$ and red arrows denote interpolation by a factor of two. $\doubleplus$~denotes concatenation along the channel dimension.}
    \end{subfigure}
    \vskip2\baselineskip
    \begin{subfigure}[t]{\textwidth}
        \centering
        \includegraphics[width=0.85\textwidth]{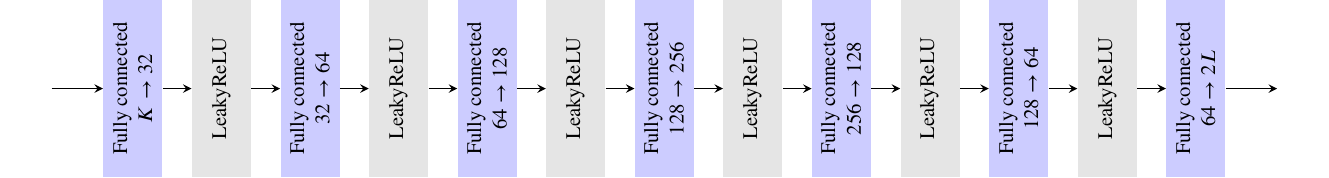}
        \caption{Temporal weights generator $\mathcal{G}_{\vec{\zeta}}$.}
    \end{subfigure}
    \vskip2\baselineskip
    \begin{subfigure}[t]{\textwidth}
        \centering
        \includegraphics[width=0.85\textwidth]{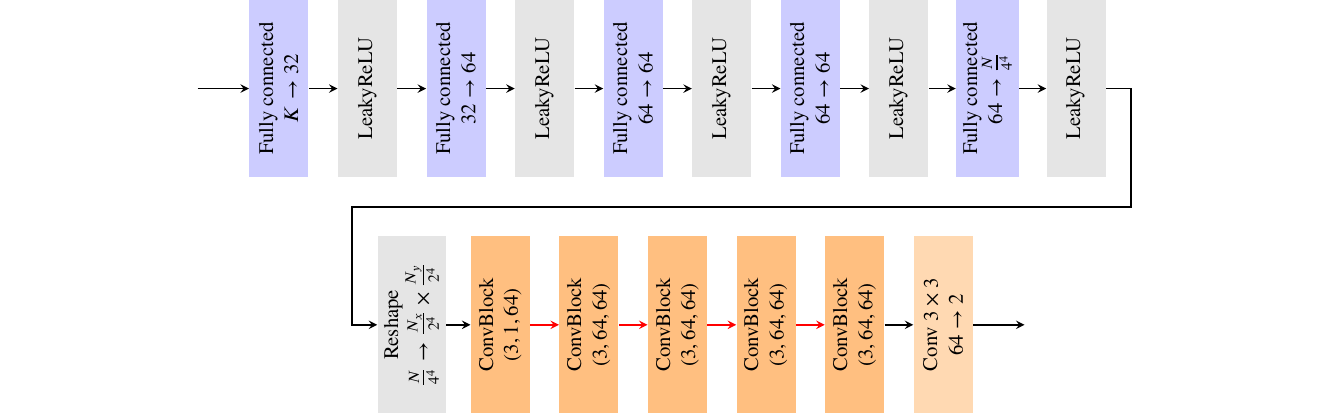}
        \caption{Deformation field generator $\mathcal{G}_{\vec{\psi}}$. $N_x$ and $N_y$ are the number of image pixels along the spatial dimensions.}
    \end{subfigure}
    \caption{Detailed network architectures of (a)~convolutional blocks, (b)~spatial dictionary generator $\mathcal{G}_{\vec{\theta}}$, (c)~temporal weights generator $\mathcal{G}_{\vec{\zeta}}$, and (d)~deformation field generator $\mathcal{G}_{\vec{\psi}}$.}
    \label{fig:architecture_details}
\end{figure}

\begin{figure*}[!t]
    \centering
    \begin{subfigure}[t]{0.23\textwidth}
        \centering
        \includegraphics[width=\textwidth]{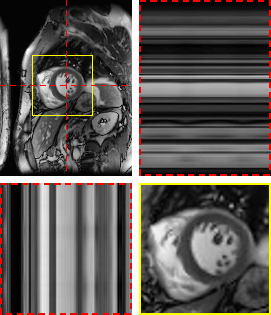}
        \caption{Real-time cine}
    \end{subfigure}
    \hspace{2mm}
    \begin{subfigure}[t]{0.23\textwidth}
        \centering
        \includegraphics[width=\textwidth]{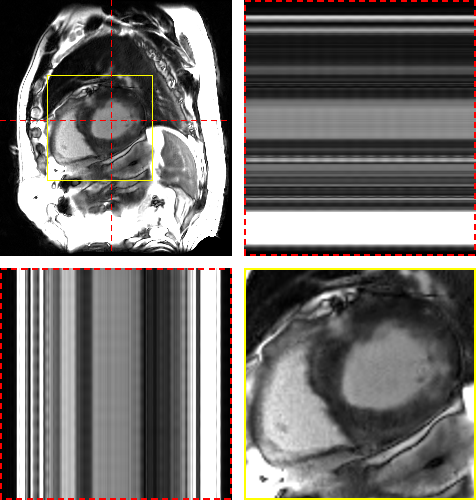}
        \caption{Free-breathing single-shot LGE}
    \end{subfigure}
    \hspace{2mm}
    \begin{subfigure}[t]{0.23\textwidth}
        \centering
        \includegraphics[width=\textwidth]{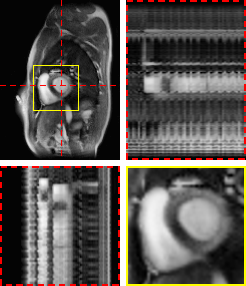}
        \caption{First-pass perfusion}
    \end{subfigure}
    \caption{Motion-compensated M-DIP reconstructions.}
    \label{fig:moco}
\end{figure*}

\end{document}